\documentstyle[aps,graphicx,multicol,amsmath]{revtex}

\begin{document}

\date{July 17, 2000}

\title{Quantum-information processing in semiconductor quantum dots}

\author{Filippo Troiani,$^*$ Ulrich Hohenester, and Elisa Molinari}

\address{
Istituto Nazionale per la Fisica della Materia (INFM) and
Dipartimento di Fisica\\
Universit\`a di Modena e Reggio Emilia, 
Via Campi 213/A, 41100 Modena, Italy}

\maketitle

\begin{abstract}

We propose an all-optical implementation of quantum-information
processing in semiconductor quantum dots, where electron-hole
excitations (excitons) serve as the computational degrees of freedom
(qubits). The strong dot confinement leads to a strong renormalization
of excitonic states, which, in analogy to NMR-based implementations of
quantum-information processing, can be exploited for performing
conditional and unconditional qubit operations.

\parindent 0pt\smallskip
[Proc. {\sc qd2000} conference; 
phys. stat. sol. (b) {\bf 224}, 849--853 (2001)]

\end{abstract}

%%%%%%%%%%%%%%%%%%%%%%%%%%%%%%%%
%%%%%%%  Introduction  %%%%%%%%%
%%%%%%%%%%%%%%%%%%%%%%%%%%%%%%%%

\begin{multicols}{2}

Quantum information, quantum computation, quantum cryptography, and
quantum teleportation represent exciting new arenas which exploit
intrinsic quantum mechanical properties [1--3].  Basic elements to
process quantum information are quantum bits (qubits), which, in
analogy to classical bits, are defined as suitably chosen two-level
systems. Much of the present excitement about quantum-information
processing (QIP) originates from the seminal discoveries of Shor and
others \cite{shor:94}, who showed that ---provided QIP can be
successfully implemented for $\sim$100--1000 qubits--- quantum
algorithms can perform some hard computations much faster than
classical algorithms, and can allow the reduction of exponentially
complex problems to polynomial complexity.

It is somewhat surprising that only a few basic requirements are needed
for a successful implementation of QIP, which, according to DiVincenzo
\cite{divincenzo:00}, can be summarized in the following five points:
{\em(i)}\/ A scalable physical system with well characterized qubits;
{\em(ii)}\/ the ability to initialize the state of the qubits;
{\em(iii)}\/ long relevant decoherence times, much longer than typical
qubit-manipulation times; {\em(iv)}\/ a ``universal'' set of quantum
gates; and {\em(v)}\/ a qubit-specific measurement capability.
Apparently, the main difficulty for a successful implementation of QIP
in a ``real physical system'' concerns the unavoidable coupling of
qubits to their environment, which leads to the process of decoherence
where some qubit or qubits of the computation become entangled with the
environment, thus in effect ``collapsing'' the state of the quantum
computer.

In this respect, semiconductor quantum dots (QDs)
\cite{hawrylak.bimberg:98} are particularly promising candidates for a
successful solid-state implementation of QIP, since in these nanoscopic
structures carriers are strongly confined in all three space
directions, leading in turn to a strongly suppressed environment
coupling. For such dots we have recently proposed an all-optical
implementation of QIP, where electron-hole excitations (excitons) serve
as the computational degrees of freedom; quantum gates can be
implemented by use of ultrashort laser pulses and coherent-carrier
control \cite{troiani:00}. Within the proposed scheme conditional qubit
manipulations, which form a cornerstone for any implementation of QIP,
naturally arise from the strong internal Coulomb interactions between
electrons and holes.  Indeed, optical single-dot spectroscopy
\cite{single-dot} has recently revealed the importance of such
Coulomb-induced renormalizations of exciton states in QDs (due to the
strong quantum confinement) [8,9].  The primary goal of the present
contribution is to highlight the similarities of our proposed scheme
with nuclear-magnetic-resonance (NMR) based implementations of QIP
\cite{nmr} and to provide a unified theoretical framework.

%%%%%%%%%%%%%%%%%%%%%%%%%%%%%%%%
%%%%%%%%%%  Theory  %%%%%%%%%%%%
%%%%%%%%%%%%%%%%%%%%%%%%%%%%%%%%

The essence of our proposal is summarized in Fig.~1 (for a discussion
of our detailed calculation see below). Fig.~1{\em(a)}\/ shows the
absorption spectrum of an empty dot (i.e., no electrons and holes
present): Two pronounced absorption peaks $X_0$ and $X_1$ can be
identified whose energy splitting is of the order of the dot
confinement (here $\sim$25 meV). As the central step within our
proposal we (tentatively) ascribe the different exciton states to the
computational degrees of freedom (qubits). Thus, for the specific case
of Fig.~1 we have two qubits ($X_0$ and $X_1$), which have value one if
exciton $x$ is populated and zero otherwise ($x=X_0,X_1$).

Let us next consider the case where the first qubit (exciton $X_0$) is
set equal to one (is populated). To simplify our analysis, we use the
fact that in most semiconductors electron-hole pairs with given spin
orientation can be selectively created by photons with a well-defined
circular polarization. Throughout this paper, we shall only consider
excitons with parallel spin orientations because of their strongly
reduced available phase space and the resulting simplified optical
density of states. For such polarizations, Fig.~1{\em(b)}\/ reports the
corresponding absorption spectrum: Due to state filling, the character
of transition $X_0$ changes from absorption to gain (i.e., negative
absorption); in addition, the higher-energetic transition is shifted to
lower energy, which is attributed to the formation of a biexcitonic
state $B$ whose energy is reduced by an amount of $\Delta$ because of
exchange interactions between the two electrons and holes, respectively
\cite{hawrylak:99}.

We next formalize the above findings. To this end, we introduce: The
exciton-operators $I_x^+$ ($I_x^-$) which create (annihilate) exciton
$x$; $I_x^z$ which has eigenvalue $1$ ($0$) if exciton $x$ is populated
(not populated). With these operators, our computational space is then
given by: The vacuum state $|0\rangle$; the single-exciton states
(where only one qubit is equal to one) $|x\rangle=I_x^+|0\rangle$; and
$|11\rangle=I_{X_0}^+I_{X_1}^+|0\rangle$, the state where both qubits
are equal to one (note that, quite generally, it is not obvious that
the biexciton state $B$ corresponds to the state $|11\rangle$ since
Coulomb interactions can mix different single-exciton states $x$;
however, as will be shown below at the example of a prototypical dot
confinement, in the strong confinement regime $|B\rangle$ is almost
parallel to $|11\rangle$, thus providing the product-type Hilbert space
required for QIP). Then, the Hamiltonian describing the exciton
dynamics in a single dot can be written (in analogy to NMR-based QIP
implementations \cite{nmr}):

\begin{equation}
  {\bf H}=\sum_{x=X_0,X_1} E_x I_x^z -
  \Delta I_{X_0}^z I_{X_1}^z,
\end{equation}

\noindent with $E_x$ the single-exciton energies. From related NMR work
it is well known that a Hamiltonian of this form can account for the
conditional and unconditional qubit operations required for QIP.

\begin{figure}
  \centerline{\includegraphics[width=0.75\columnwidth]{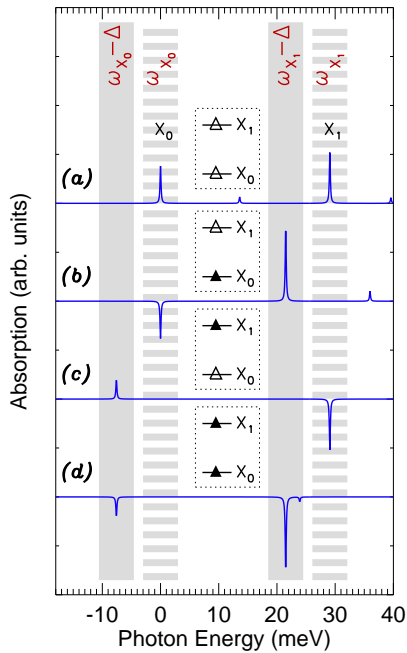}}
  \caption{
  Absorption spectra for semiconductor quantum dot described in the
  text, which is initially prepared in (see insets): {\em(a)}\/ vacuum
  state; {\em(b)}\/ exciton $|X_0\rangle$ state (exciton groundstate);
  {\em(c)}\/ exciton $|X_1\rangle$ state; {\em(d)}\/ biexciton
  $|B\rangle$ state. Photon energy zero is given by the groundstate
  exciton $X_0$.
  }
\end{figure}

%%%%%%%%%%%%%%%%%%%%%%%%%%%%%%%%
%%%%%%%%%%  Results  %%%%%%%%%%%
%%%%%%%%%%%%%%%%%%%%%%%%%%%%%%%%

As a representative example for the Hamiltonian (1), in this paper we
consider a prototypical dot confinement that is parabolic in the
$(x,y)$-plane and box-like along $z$ \cite{dot}. Following the
procedure outlined in Ref.~\cite{troiani:00}, we expand exciton and
biexciton states within the subspaces of spin-selective electron-hole
excitations, and obtain the excitonic eigenenergies and eigenstates
through direct diagonalization of the two- and four-particle
Schr\"odinger equations (accounting for all possible carrier-carrier
Coulomb interactions). The single-exciton operators can then be
expressed through $I_x^+=\sum_{\mu\nu}\Psi_{\mu\nu}^x c_\mu^\dagger
d_\nu^\dagger$, where $\Psi_{\mu\nu}^x$ is the exciton wavefunction and
$c_\mu^\dagger$ ($d_\nu^\dagger$) creates an electron in state $\mu$
(hole in state $\nu$). We checked that state $|11\rangle$ has more than
95\% overlap with the true Coulomb-renormalized biexciton state $B$,
which finally justifies the use of the Hamiltonian (1).

\begin{figure}
  \centerline{\includegraphics[width=0.75\columnwidth]{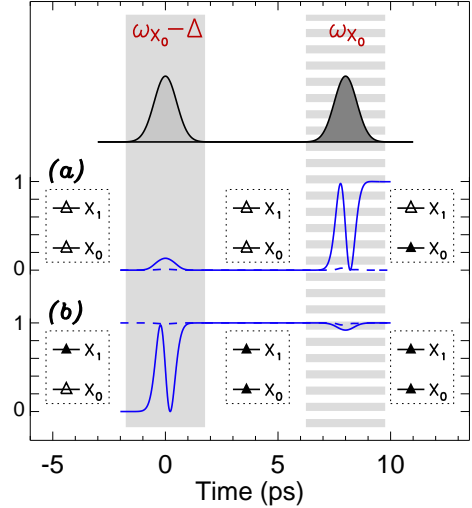}}
  \caption{ 
  Results of our simulations (neglecting dephasing) of qubit
  manipulations by means of coherent-carrier control: {\em(a)}\/
  initial state $|00\rangle$; {\em(b)}\/ initial state $|01\rangle$.
  The solid (dashed) lines correspond to the transient population
  $\langle I_{X_0}^z\rangle_t$ of exciton $X_0$ ($\langle
  I_{X_1}^z\rangle_t$).  The sequence of pulses and the corresponding
  photon energies are indicated at the top of the figure (see also
  insets); for the envelopes of the laser pulses we use Gaussians
  $\propto\exp(-t^2/2\tau^2)$, with $\tau=0.5$ ps. The first operation
  at time zero corresponds to a conditional operation where the second
  qubit acts as the control-qubit; the sequence of the two pulses
  corresponds to a NOT operation on the first qubit.
  }
\end{figure}

To understand how quantum gates can be implemented within our scheme,
we first observe in Fig.~1 that the appearance and disappearance of
peaks at the frequencies indicated by the (solid and dashed) shaded
areas {\em conditionally}\/ depends on the setting of specific qubits:
E.g., the optical transitions at $\omega_{X_0}-\Delta$ is {\em only}\/
present if the second qubit is set equal to one (Figs.~1{\em(c)}\/ and
1{\em(d)}), whereas the transition at $\omega_{X_0}$ {\em only}\/
appears if the second qubit is set equal to zero (Figs.~1{\em(a)} and
1{\em(b)}); an analogous behaviour can be found for the other two
transitions. Indeed, it is this conditional on- and off-switching of
optical transitions that enables $\pi$-laserpulses to modify the state
of one qubit or not, depending on the setting of the other qubit. For
instance, as shown in Fig.~2, a $\pi$-pulse at frequency
$\omega_{X_0}-\Delta$ changes the exciton population of $X_0$ only if
exciton $X_1$ is populated: Such a transformation corresponds to a
controlled-NOT (C-NOT) operation in which the second qubit (exciton
$X_1$) acts as a control qubit. In addition, a combination of such a
pulse with a $\pi$-pulse at frequency $\omega_{X_0}$ changes the state
of exciton $X_0$ (target qubit) independently of the setting of the
second qubit (NOT operation; see sequence of the two pulses in
Fig.~2).

%%%%%%%%%%%%%%%%%%%%%%%%%%%%%%%%
%%%%%%%%%  Conclusion  %%%%%%%%%
%%%%%%%%%%%%%%%%%%%%%%%%%%%%%%%%

Let us finally comment to which extent DiVicenzo's five requirements
for QIP are fulfilled within our scheme: {\em(i)}\/ Although in this
contribution we have only discussed a two-qubit implementation of QIP,
one can expect that analogous schemes in arrays of coupled QDs (which
arise naturally in dot fabrication) could allow implementation of a
moderate number of qubits (see also \cite{rossi:00}); {\em(ii)}\/
because of the nano-second electron-hole recombination times in
semiconductors, ``initialization'' is not expected to cause major
problems; {\em(ii)}\/ within our scheme, relevant decoherence times are
of the order of several tens of picoseconds \cite{time-dependent} which
is not too much longer than the estimated sub-picosecond
qubit-manipulation times \cite{troiani:00}; use of excitonic
groundstates in an array of QDs could help to improve such performance,
because there decoherence times are of the order of nano-seconds;
{\em(iv)}\/ as shown in Figs.~1 and 2 conditional and unconditional
qubit operations are implementable; {\em(v)}\/ finally, qubit-specific
measurements could be performed in analogy to trapped-ion
implementations of QIP \cite{cirac.zoller:95}, although more detailed
strategies still have to be worked out. In conclusion, we expect the
present proposal to be particularly promising for the first successful
demonstration of QIP in this class of materials, and to evaluate
whether optical excitations in QDs indeed can serve as reliable
qubits.

We are grateful to F. Rossi and P.  Zanardi for
most helpful discussions. This work was supported in part by the EU
under the TMR Network "Ultrafast" and the IST Project SQID, and by INFM
through grant PRA-SSQI.

\end{multicols}
\begin{multicols}{2}

\end{multicols}


\begin{thebibliography}{10}

\bibitem[*]{*} E-mail: troiani@unimo.it.

\bibitem{steane:97}
A. Steane. Rep. Prog. Phys. {\bf 61}, 117 (1998).

\bibitem{physics.world:98}
See, e.g., the articles in the march 1998 issue of Phys. World.

\bibitem{divincenzo:00}
D. P. DiVincenzo, quant-ph/0002077.

\bibitem{shor:94}
P. Shor, {\em Proceedings of the 35th Annual Symposium on the Foundations
of Computer Science}\/ (IEEE Press, Los Alamitos, 1994), p. 124.

\bibitem{hawrylak.bimberg:98}
L. Jacak, P. Hawrylak, and A. Wojs, 
{\em Quantum Dots}\/ (Springer, Berlin, 1998);
D. Bimberg, M. Grundmann, and N. Ledentsov, 
{\em Quantum Dot Heterostructures}\/ (John Wiley, New York, 1998).

\bibitem{troiani:00}
F. Troiani, U. Hohenester, and E. Molinari,
Phys. Rev. B {\bf 62}, R2263 (2000).

\bibitem{single-dot}
J.~Motohisa, J.J.~Baumberg, A.P.~Heberle, J.~Allam,
Solid-State Electronics {\bf 42}, 1335 (1998);
L.~Landin, M.S.~Miller, M.E.~Pistol, C.E.~Pryor, L.~Samuelson,
Science {\bf 280} 262 (1998);
E.~Dekel, D.~Gershoni, E.~Ehrenfreund, D.~Spektor,
J.M.~Garcia, M.~Petroff,
Phys. Rev. Lett. {\bf 80}, 4991 (1998);
A. Zrenner et al., Physica B {\bf 256--258}, 300 (1998).

\bibitem{hohenester:99}
U. Hohenester, F. Rossi, and E. Molinari,
Solid-State Commun. {\bf 111}, 187 (1999).
     
\bibitem{hawrylak:99}
P. Hawrylak, Phys. Rev. B {\bf 60}, 5597 (1999).

\bibitem{nmr}
N.A. Gershenfeld and I.L. Chuang, Science {\bf 275}, 350 (1997);
D. G. Cory {\em et al.},\/ quant-ph/0004104.

\bibitem{dot} In our cylindrical QD the confinement energies due to
the in-plane parabolic potential are $\omega_o^{(e)}=20$ meV for
electrons, and $\omega_o^{(h)}=3.5$ meV for holes; with this choice,
electron and hole wavefunctions have the same lateral extension. The
quantum-well confinement along $z$ is such that the intersubband
splittings are much larger than $\omega_o^{(e,h)}$. Material parameters
for GaAs are used. In our calculations we keep for electrons and holes,
respectively, the ten single-particle states of lowest energy.

\bibitem{time-dependent}
N.H. Bonadeo, J. Erland, D. Gammon, D.S. Katzer, D. Park, and D.G. Steel,
Science {\bf 282}, 1473 (1998);
Y. Toda, T. Sugimoto, M. Nishioka, and Y. Arakawa, 
Appl. Phys. Lett. {\bf 76}, 3887 (2000).

\bibitem{rossi:00}
E. Biolatti, R. C. Iotti, P. Zanardi, and Fausto Rossi,
Phys. Rev. Lett. {\bf 85}, 5647 (2000).

\bibitem{cirac.zoller:95}
J. I. Cirac and P. Zoller, Phys. Rev. Lett. {\bf 74}, 4091 (1995).


\end{thebibliography}
\end{document}